# Multifunctional 2D CuSe monolayer nanodevice

Yipeng An[1,2,3], Yusheng Hou[2], Hui Wang[2], Jie Li[2], Ruqian Wu[2,3], Chengyan Liu[2], Tianxing Wang[1] and Jutao Jiao[1]

[1]College of Physics and Materials Science & International United Henan Key Laboratory of Boron Chemistry and Advanced Energy Materials, Henan Normal University, Xinxiang 453007, China

[2]Department of Physics and Astronomy, University of California, Irvine, California 92697, USA

[3]Author to whom any correspondence should be addressed.

**E-mail:** ypan@htu.edu.cn(Y. An);wur@uci.edu(R. Wu)

**Keywords:** Nanodevices, Electronic transport, CuSe monolayer, negative differential resistance, photon devices

**Abstract:**

In a very recent experimental work [Gao et al., **2018** *Adv. Mater.* **30**, 1707055], a graphene-like CuSe monolayer (ML) was realized. Motivated by this success, we performed first-principles calculations to investigate its electronic transport and photoelectronic properties. We find that the CuSe ML shows a strong electrical anisotropy, and its current-voltage (*I–V*) curves along the zigzag and armchair directions are noticeably different. The CuSe ML also displays a useful negative differential resistance (NDR) effect along the both directions when the bias is beyond 1.0 V. Moreover, it has a large photon absorption to orange light. Our study suggests that CuSe ML is a multifunctional material and has various potential applications in electrical-anisotropy-based, NDR-based, and even optical nanodevices.

# 1. Introduction

Two-dimensional (2D) materials have attracted an extensive research interest since the successful isolation of graphene by mechanical exfoliation from graphite in 2004 [1]. Many 2D layered materials have been predicted theoretically, and a few of them have been realized experimentally, including silicene [2-4], hexagonal boron nitride (h-BN) [5, 6], transition metal dichalcogenides (TMDs) [7, 8], phosphorene [9, 10], MXene [11, 12], borophene [13-16], and their derivatives [17-19]. 2D MLs typically show distinguished mechanical, electronic, optical, and thermal conduction properties, and are hence promising for the design of novel nanodevices. A variety of monolayer-based prototype devices has been proposed [20-24] and fabricated [9, 25, 26], and they demonstrated a great potential for technological innovations, particularly for the next-generation electronic and photoelectronic applications.

In a very recent experiment [27], a graphene-like transition metal monochalcogenide CuSe ML was successfully synthesized on Cu(111) substrate by means of molecular beam epitaxy (MBE). High resolution scanning tunneling microscopy (STM), angle resolved photoemission spectra (ARPES), and first-principles calculations have been applied for the characterization and determination of its honeycomb structure and electronic properties. It is interesting that the honeycomb CuSe ML possesses two 2D Dirac nodal line fermions (DNLFs), protected by the mirror reflection symmetry (MRS). To exploit CuSe ML in nanodevices, it is imperative to investigate this new 2D material more comprehensively. For example, its electronic transport and photoelectronic properties have not been examined, neither theoretically or experimentally.

In the present work, we carry out a systematic research on the electronic, transport and optical properties of the 2D CuSe ML (see figure 1(a)) by means of first-principles calculations. Our results show that 2D CuSe has a remarkable electrical anisotropy, namely, its *I–V* curves along the zigzag and armchair directions are noticeably different. In particular, the CuSe ML shows strong negative differential resistances along both directions. Furthermore, it displays a strong photo response to orange light. Our findings suggest that the CuSe-based ML as a new multifunctional material can be used for a variety of purposes.

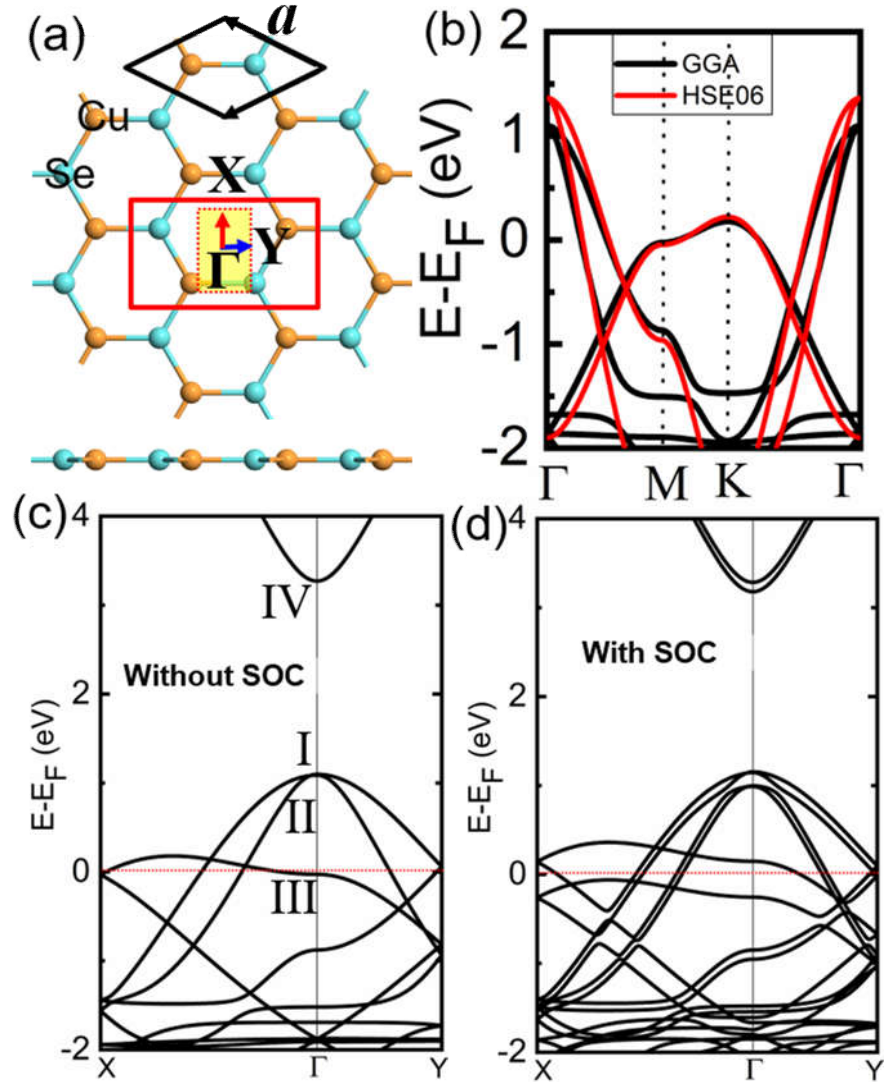

**Figure 1.** (a) Atomic structure of CuSe monolayer. The black/red box refers to its hexagonal (H)/simple orthorhombic (SO) unit cell. The first Brillouin zone of SO unit cell is embedded in its box. (b) Band structures of hexagonal CuSe ML. Bands of CuSe along $X$–$\Gamma$–$Y$ path without (c) and with (d) SOC. I-IV in (c) refer to the four bands near the Fermi level which is set to zero. I and II are degenerated around $\Gamma$ point.

## 2. Method

The calculations of electronic transport and photoelectronic properties of 2D CuSe ML are performed using the density functional theory and nonequilibrium Green's function approach as implemented in the ATK-DFT software [28-30]. The Perdew-Burke-Ernzerhof (PBE) scheme of the generalized gradient approximation (GGA) [31, 32] is adopted to describe the exchange-correlation effect among electrons. The core electrons of all atoms are described by the optimized norm conserving Vanderbilt (ONCV) pseudopotentials [33], and wave-functions of valence states are expanded as linear combinations of atomic orbitals (LCAO) with the SG15 basis set, which are fully relativistic and can provide comparable results to the all-electron method. The cutoff for the mesh density of basis expansion is set as 200 Ry. The atomic structures are fully optimized until the residual force on each atom is smaller than 0.01 eV/Å and the total energy tolerance is below $10^{-6}$ eV, respectively. Higher-level HSE06 method is used to confirm its band properties [34, 35]. For the transport calculations, we use a 1×9×100 Monkhorst-Pack **k** points grid to sample the Brillouin zone of the electrodes. The spin-orbital coupling correction is also taken into account as a comparison.

## 3. Results and discussion

Figure 1(a) shows the optimized atomic configuration of the 2D CuSe ML. The honeycomb lattice parameter *a* is 3.98 Å, and the corresponding Cu-Se bond length is 2.30 Å. The band structure for a primitive unit cell is plotted in figure 1(b), which shows a metallic character, in good agreement with the previous report [27]. The result of higher-level HSE06 method gives the consistent band structure and confirms its metallic character (see figure 1(b)). In general, 2D honeycomb structures have different mechanical, electronic, and transport properties along the zigzag and armchair directions. For instance, graphene shows a metallic character along the zigzag direction but has a semiconducting feature along the armchair one [36]. To better show the anisotropic characteristics of the CuSe ML, we plot its bands along the $X$–$\Gamma$–$Y$ path for a rectangular unit cell, i.e., the zigzag direction ($\Gamma$–$X$) and armchair one ($\Gamma$–$Y$) in figure 1(c). Although CuSe ML appears to be metallic in both directions, the difference is also obvious as more bands cross the Fermi level ($E_F$) along the zigzag direction than along the armchair one. Figure 1(d) shows the band structures with spin-orbit coupling (SOC), which only causes tiny band splits and would not obviously influence its electronic transport properties.

To get a more quantitative description of this difference, we construct a two-probe structure of 2D CuSe ML (see figure 2(a)) and directly calculate its electric conductivities along the zigzag (X axis) and armchair (Y axis) directions, labeled as z-CuSe and a-CuSe, respectively. The two-probe structure has a periodicity perpendicular to transport direction (i.e., from the drain (D) to the source (S) electrode), and its third orthogonal direction is a slab with a 15 Å thick vacuum along the normal direction of CuSe ML plane. Both the drain and source electrodes are represented by a large supercell that is semi-infinite in length along the transport direction. Under a bias $V_b$, the current $I$ is obtained by using the Landauer–Büttiker approach [37]

$$I(V_b) = \frac{2e}{h} \int_{-\infty}^{\infty} T(E, V_b)[f_D(E - \mu_D) - f_S(E - \mu_S)] dE . \quad (1)$$

Here, $T(E, V_b)$ is the bias-dependent transmission coefficient, determined from the Green's functions; $f_{D/S}$ are the Fermi-Dirac distribution functions of the drain/source electrodes; their electrochemical potentials are shifted as $\mu_D$ (= $E_F$ − $eV_b/2$) and $\mu_S$ (= $E_F$ + $eV_b/2$), respectively. Obviously, [$\mu_D$, $\mu_S$] sets a bias window (BW) for electron transport and we will focus on states in this energy range in following discussions. More details about this approach can be seen in the previous reports [28-30, 38, 39].

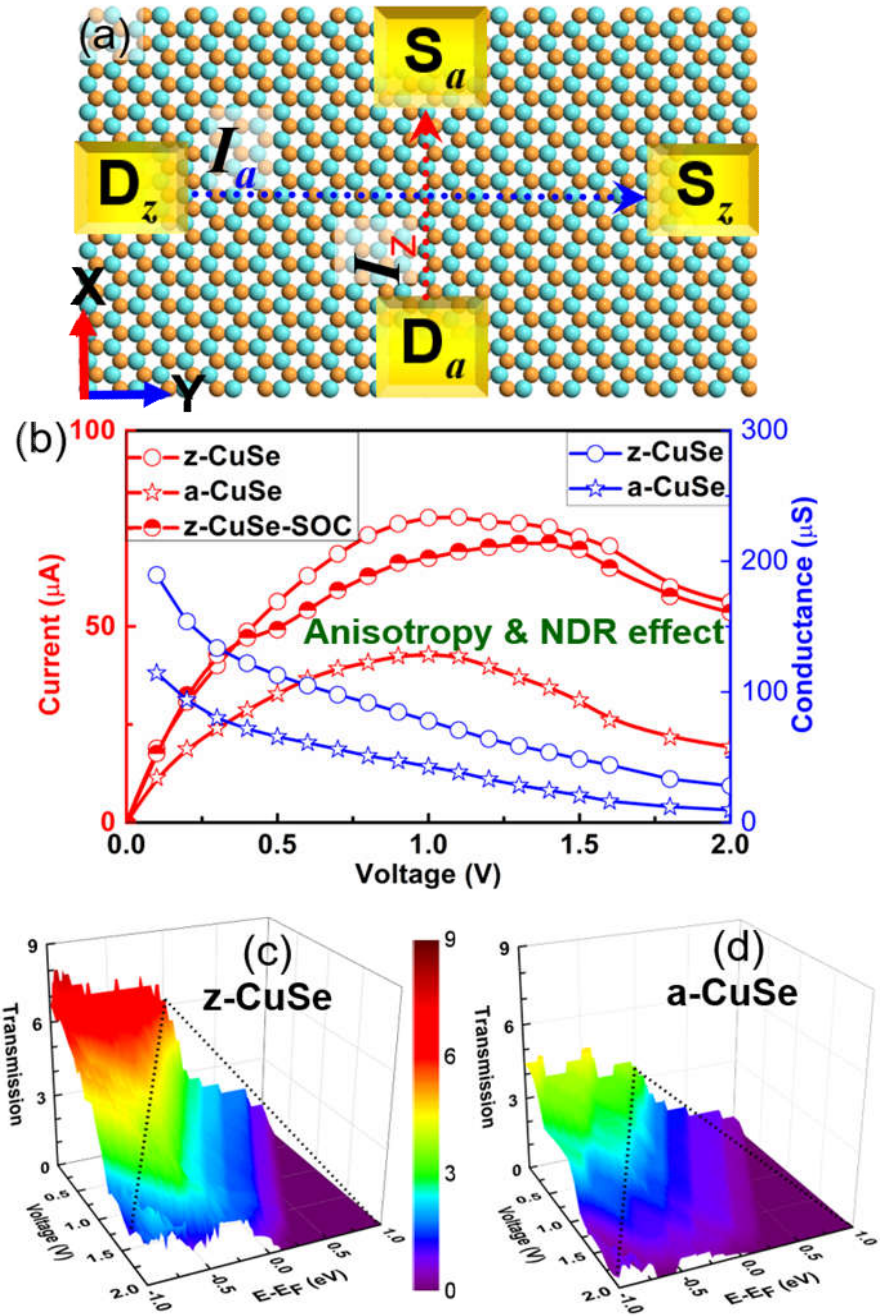

**Figure 2.** (a)Schematic of CuSe nanodevice. $D_{z/a}$ and $S_{z/a}$ refer to the drain and source electrodes along the zigzag (X axis)/armchair (Y axis) direction, respectively. $I_{z/a}$ refers to the current through z-CuSe/a-CuSe diode. (b) *I–V* and *I–G* curves of z-CuSe and a-CuSe diodes. Transmission spectra under various biases of z-CuSe (c) and a-CuSe (d). The Fermi level is set to zero.

Figure 2(b) shows the *I–V* curves of z-CuSe and a-CuSe two-probe systems, which are noticeably different. The CuSe ML has better conductivity along the zigzag direction than along the armchair one. Its ratio of current anisotropy $\eta=I_z/I_a$ ($I_z$ and $I_a$ refer to the currents of z-CuSe and a-CuSe, respectively) is about 2, larger than that of most 2D materials [40, 41]. Furthermore, it shows a prominent negative differential resistance effect for both the z-CuSe and a-CuSe when the bias around 1.0 V. Note that NDR is an important feature of electronic materials and it is useful for many applications such as memory cells [42], threshold logic [43], electronic oscillators [44], amplifiers [45], and particularly at microwave frequencies [46]. In additions, this effect is often characterized by two important factors that are obviously dependent on the materials and can change significantly in experiments. The first factor is the NDR threshold voltage (NDR-TV) where the current reaches its maximum. The second one is the alleged peak-to-valley ratio (PVR) between the maximal (peak) and minimal (valley) currents. Generally, it is desired to have materials with low NDR-TV for minimizing the power consumption along with large PVR for maximizing the performance [47]. To this end, both z-CuSe and a-CuSe have the same NDR-TV as small as 1.0 V, better than most materials reported so far [48], and a large PVR because the current decrease quite rapidly under a high bias. Therefore, we may propose that a 2D CuSe ML is an excellent NDR material, much better than $ZrB_2$ ML [41] and thiol-terminated Ru(∥) bis-terpyridine molecular junction proposed before [49]. What's more, the conductance (G) of CuSe ML along the zigzag or armchair direction is gradually decreased as the bias increases and will converge to almost zero under a specific high bias (see the *I–G* curves shown in figure 2(b)). Note that the SOC does not obviously influence the electronic transport of CuSe ML due to its tiny band splits.

For instance, the *I–V* curve of z-CuSe with SOC (labeled as z-CuSe-SOC) is almost consistent with that without SOC (see figure 2(b)).

To understand the electrical anisotropy of CuSe ML and NDR effect of CuSe ML, we first examine the transmission spectra of z-CuSe and a-CuSe under the various biases (see figures 2(c) and 2(d)). The CuSe ML, especially for the z-CuSe, has better electron transmission channels under the low bias (i.e., below 1.0 V), which gives rise to upward current as the bias increases. However, the electron transmission coefficients drop drastically after that (i.e., beyond 1.0 V), even to zero near the upper edge of the bias window. This results in the strong anisotropy and NDR phenomena, as the current is the integral of transmission coefficients over the bias window (by equation 1).

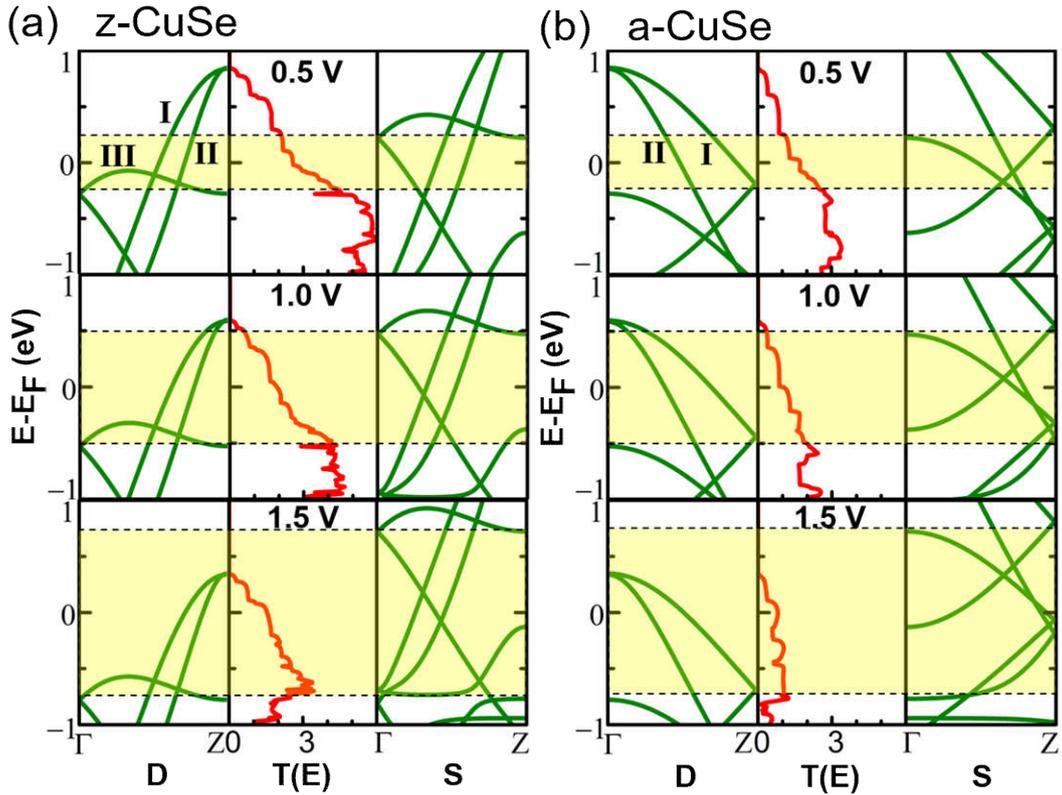

**Figure 3.** Transmission spectra and band structures for the drain and source electrodes of the z-CuSe (a) and a-CuSe (b) under the biases of 0.5, 1.0, and 1.5 V, respectively. The Fermi level is set to zero.

Basically, the electronic transport of 2D monolayers is mostly dominated by their band structures, from the inter- and intra-band transitions around the $E_F$. Figure 3 shows the band structures and transmission spectra of the CuSe ML under biases of 0.5, 1.0, and 1.5 V, respectively. As a forward bias is applied to the drain and source electrode, their bands shift down and up accordingly. The transmission coefficients are mainly determined by the band overlap between the drain and the source electrodes. For z-CuSe, the transmission coefficients are mostly related to the bands of drain electrode (see figure 3(a)), and only its bands **I**, **II**, and **III** within the bias window give rise to effective contributions to the current. As bias increases (from 0 to 1.0 V), the expansion of the bias window results in the upward *I–V* curve, even though the transmission coefficient slightly decreases. However, as the bias increases further (such as 1.5 V), the bias window hits a gap of the drain electrode and the current decreases, causing the NDR phenomenon. The same mechanism is applicable for the NDR of a-CuSe. Differently, only two bands (**I and II**) of its drain electrode contribute to the electron transmission, leading to smaller conductance.

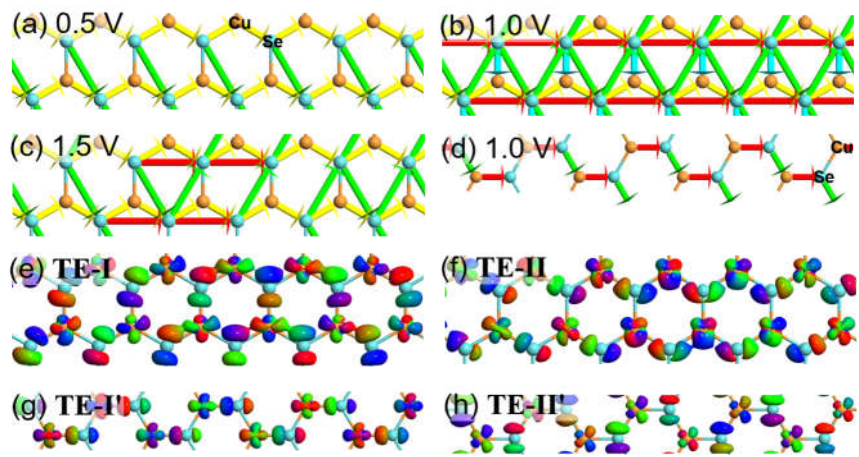

**Figure 4.** Transmission pathways of z-CuSe at the Fermi level under 0.5 (a), 1.0 (b), and 1.5 V (c), respectively. (d) Transmission pathways of a-CuSe at 1.0 V. (e) **TE-I** and (f) **TE-II** of z-CuSe, and (g) **TE-I'** and (h) **TE-II'** of a-CuSe at the $E_F$ under 1.0 V, respectively.

We further analyze the electron transmission pathways (i.e., local current) [50] of z-CuSe and a-CuSe. This splits the total transmission coefficient into local bond contributions $T_{ij}$. The pathways across the boundary between two parts (A and B) give rise to the total transmission coefficient

$$T(E)=\sum_{i\in A, j\in B} T_{ij}(E) . \qquad (2)$$

In general, there are two types of local current pathways: (i) bond current (i.e., *via* chemical bonds between two neighboring atoms), and (ii) hopping current (i.e., *via* electron hopping between atoms on the same lattice) [51]. At a low bias (see figure 4(a) at 0.5 V), there are two different local current pathways for z-CuSe: (i) Cu−Se−Cu bond current (yellow arrows) and (ii) Se→Se hopping current (green arrows). As the bias reaches up to 1.0 V (see figure 4(b)), more current pathways appear, e.g., (iii) Se→Se hopping current (red arrows) and (iv) Se−Cu bond current (cyan arrows). Note that the fourth Se−Cu bond current pathway is perpendicular to the transport direction and hence gives little contribution to the total current. Under a high bias (such as 1.5 V), the current pathways (see figure 4(c)) decreases. For a-CuSe, only one type of current pathways, namely, step-like Cu−Se−Cu−Se bond current pathways (see figure 4(d)), contributes to electron transport.

Obviously, both Cu and Se orbitals contribute to electron transmission of z-CuSe and a-CuSe as eigenstates near the $E_F$ under results from the Cu-Se hybridization (see figures 4(e) to 4(h)). Both z-CuSe and a-CuSe have the double-degenerate transmission eigenstates at the $E_F$ under the bias of 1.0 V, namely, **TE-I** and **TE-II** for z-CuSe, **TE-I'** and **TE-II'** for a-CuSe, respectively. These states become localized or even empty under a high bias, leading to the NDR effect.

Another important factor for the design of electrical and electronic circuits is threshold current density (CD). Most electrical conductors have a finite resistance and dissipate power as Joule heat which may cause melting of conductors. Our calculations indicate that both the CD of z-CuSe and a-CuSe first increase and then decrease as the bias is beyond 1.0 V (see figures 5(a) to 5(f)). For instance, the CD of z-CuSe increases from $4.0\times10^5$ nA/Å$^2$ at 0.5 V (see figure 5(a)) to $7.6\times10^5$ nA/Å$^2$ at 1.0 V (see figure 5(b)), then gradually decreases due to the NDR effect, and hence it is self-protected.

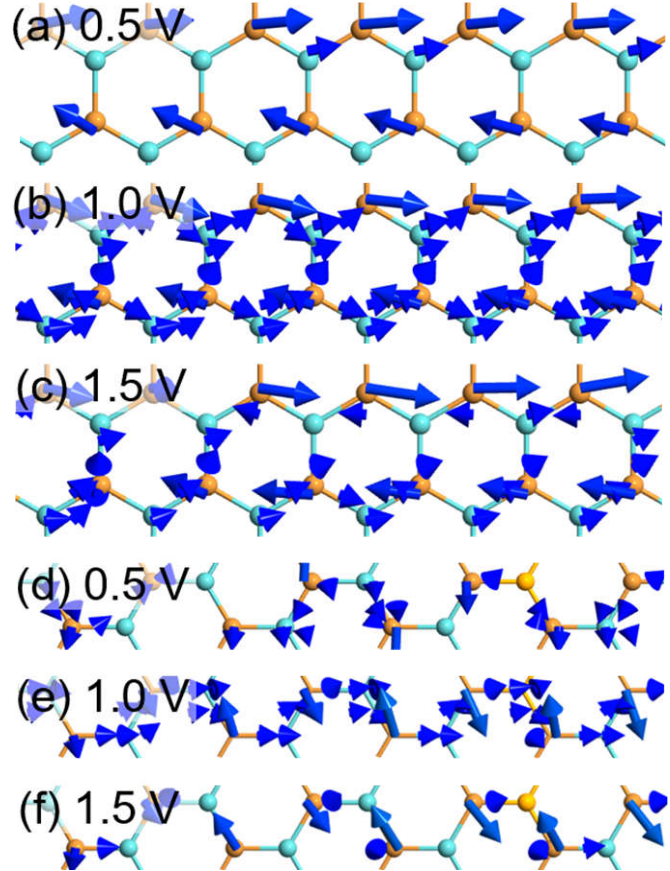

**Figure 5.** Current density of CuSe ML under 0.5, 1.0, and 1.5 V, respectively. (a)-(c) for z-CuSe, (d)-(f) for a-CuSe.

Different from the conventional photoelectronic devices, the monolayer materials may have exceptional transport and optical properties. For instance, the large-area single and few-layers of graphene ultrafast photodetectors had been prepared in experiments[52]. Now we investigate if CuSe ML has a potential for applications in photoelectronics. Figure 6(b) shows the optical absorption coefficient, $\alpha$, as a function of photon energy of the zigzag and armchair CuSe ML. As expected, the photon absorption spectra of CuSe ML along these two directions are identical. There are two absorption peaks (near the energy E = 2.0 and 0.9 eV) within the AM-1.5 photon energy range, including one in the visible light region (orange light). They result from the electron transitions from the band I to IV and from the III to II or degenerated I around the $\Gamma$ point (see figure 1(c)), respectively. The absorption coefficients at these two peaks, 1.2-1.4×$10^5$ cm$^{-1}$ are rather large. Therefore, the CuSe ML can be a promising candidate for the use of photodetectors, such as to examine the orange and infrared light.

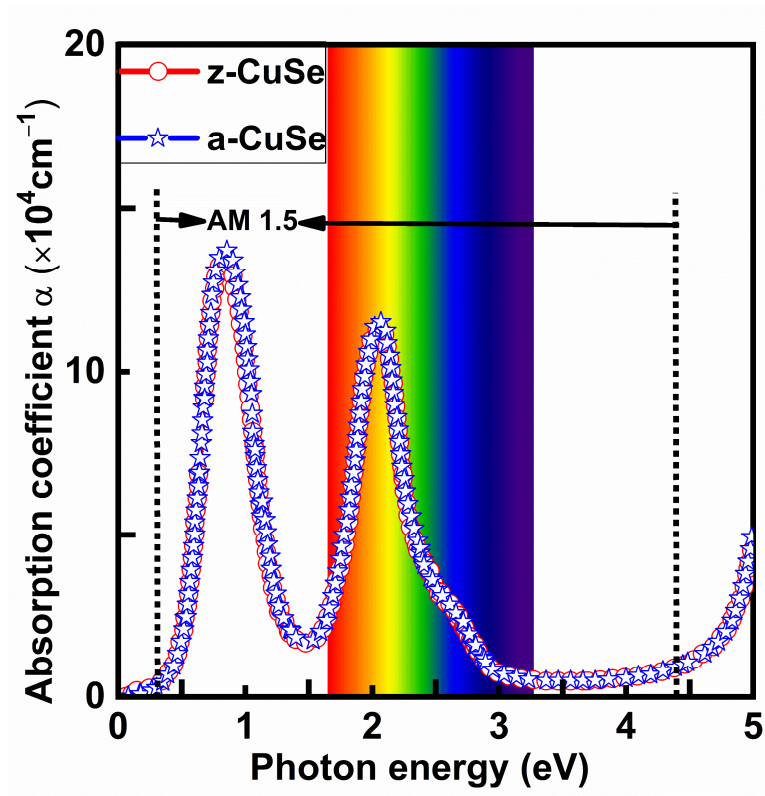

**Figure 6.** Photon absorption coefficient $\alpha$ of z-CuSe and a-CuSe. The embedded spectrum patterns indicate the visible light region.

## 4. Conclusions

In summary, we systematically study the electronic transport and photoelectronic properties of 2D CuSe ML along the zigzag and armchair directions by means of the first-principles calculations. Our results show that CuSe ML shows a strong electrical anisotropy along these two perpendicular directions, based on its diverging *I–V* curves. Both z-CuSe and a-CuSe show a significant NDR effect with low threshold voltage and large PVR. Furthermore, CuSe ML has a large photon absorption coefficient to the orange light. Our findings suggest that the multifunctional CuSe ML has a variety of applications in electrical-anisotropy-based, NDR-based, and photoelectronic nanodevices.

**Acknowledgments**

The work at the University of California at Irvine was supported by the US DOE-BES under Grant DE-FG02-05ER46237. The work at Henan Normal University was supported by the National Natural Science Foundation of China (Grant Nos. 11774079 and U1704136), the CSC (Grant No. 201708410368), the Natural Science Foundation of Henan Province (Grant No. 162300410171), the young backbone teacher training program of Henan province's higher education, the Science Foundation for the Excellent Youth Scholars of Henan Normal University (Grant No. 2016YQ05). We also thank X. Dai at Zhengzhou Normal University and H. Da at Nanjing University Posts and Telecommunications for helpful discussion, and the High-Performance Computing Centre of Henan Normal University.

**ORCID**

Yipeng An: 0000-0001-5477-4659

Ruqian Wu: 0000-0002-6156-7874

Hui Wang: 0000-0001-9972-2019

Tianxing Wang: 0000-0003-3659-8801